\documentclass{article}
\usepackage{spconf,amsmath,amsfonts,graphicx,booktabs}
\usepackage{url}
\usepackage[hidelinks]{hyperref}

\makeatletter
\let\ICASSP@thebibliography\thebibliography
\renewcommand{\thebibliography}[1]{%
  \ICASSP@thebibliography{#1}%
  \setlength{\itemsep}{0pt}%
  \setlength{\parskip}{0pt}%
  \setlength{\parsep}{0pt}%
}
\makeatother

\makeatletter
\newcommand{\leftalignedcaption}[2]{%
  \begingroup
  \long\def\@makecaption##1##2{%
    \vskip 6pt
    \noindent ##1. ##2\par}%
  \caption{#1}%
  \label{#2}%
  \endgroup}
\long\def\@makecaption#1#2{%
  \vskip 6pt
  \setbox\@tempboxa\hbox{#1. #2}%
  \ifdim \wd\@tempboxa >\hsize
    #1. #2\par
  \else
    \hbox to\hsize{\hfil\box\@tempboxa\hfil}%
  \fi}
\makeatother

\newcommand{\method}{\textsc{EditVoice}}

\title{EditVoice: Variable-Length Non-Autoregressive Zero-Shot TTS and Speech Editing with Edit Flows}

\name{\shortstack{Hongyao Deng$^{1}$ \qquad Wenhao Guan$^{2}$ \qquad Xuetao Lin$^{1}$ \qquad Peijie Chen$^{1}$ \\
Weijie Wu$^{1}$ \qquad Lin Li$^{2,*}$ \qquad Qingyang Hong$^{1,*}$}%
\thanks{$^{*}$ Corresponding author.}}
\address{$^{1}$School of Informatics, Xiamen University, China \\
$^{2}$School of Electronic Science and Engineering, Xiamen University, China}

\begin{document}
\ninept
\maketitle

\begin{abstract}
Recent non-autoregressive (NAR) zero-shot text-to-speech (TTS) models generate in parallel but typically require the target sequence length to be specified before generation. We introduce EditVoice, to our knowledge the first variable-length NAR zero-shot TTS model, which uses Edit Flows to jointly update speech content and sequence length through insertions, deletions, and substitutions.  EditVoice adopts speech-infilling training, which unifies zero-shot TTS and text-based speech editing and allows both prefix and suffix speech prompt placements at inference. We introduce Complementary Prompt Sampling (CPS) to leverage the complementary Edit Flow predictions induced by the two prompt placements. We further find that EditVoice can edit source and model-generated
speech beyond its training sources. We use this generalization for
end-to-end editing and training-free post-generation refinement. With the Edit Flow model trained on 10K h of GigaSpeech, EditVoice
demonstrates competitive zero-shot TTS performance on Seed-TTS Eval EN
and LibriSpeech-PC and speech editing performance on RealEdit. Audio samples are available at
\url{https://dhy02.github.io/editvoice-demo/}.
\end{abstract}


\section{Introduction}
\label{sec:intro}

Modern zero-shot text-to-speech (TTS) models can be broadly divided
into two paradigms: autoregressive (AR) and non-auto\-regressive (NAR). NAR TTS has explored continuous-space diffusion
and flow models~\cite{le2023voicebox,guan2024reflowtts,
eskimez2024e2tts,chen2025f5tts,zhu2025zipvoice} and discrete diffusion
and flow models~\cite{wang2024maskgct,zhu2026omnivoice,
fan2026lladatts,yin2026lunatts,nguyen2026diflowtts}. AR decoding
naturally produces variable-length sequences but commits to previous
predictions, allowing local errors to propagate~\cite{moon2026deltatts}.
NAR generation updates speech representations in parallel, but
typically operates on a target sequence whose length is specified or
predicted before generation. This also restricts iterative correction
to a fixed-length sequence, as in PALLE~\cite{yang2025palle}.
Edit Flows~\cite{havasi2025edit} model variable-length sequences
through insertion, deletion, and substitution. We introduce \method{},
a NAR zero-shot TTS model based on Edit Flows that jointly updates
speech content and sequence length during sampling.

Speech infilling provides a common formulation for zero-shot TTS and
text-based speech editing: the model reconstructs masked speech from
text and surrounding audio context~\cite{le2023voicebox}. This
formulation has been used with continuous flow models
~\cite{le2023voicebox,eskimez2024e2tts,chen2025f5tts},
autoregressive codec models~\cite{peng2024voicecraft,
wang2024ssrspeech}, and masked diffusion models
~\cite{fan2026lladatts,yin2026lunatts}. These approaches perform
editing by regenerating masked speech regions. We train \method{} with
random-span speech infilling, allowing the same model to perform
zero-shot TTS and localized text-based speech editing.

This training makes both prefix and suffix prompt placements available
for zero-shot TTS and speech editing. Joint use of the two
placements remains unexplored in prior infilling-based TTS
~\cite{eskimez2024e2tts,chen2025f5tts}. We find that the two placements
provide complementary Edit Flow predictions. We introduce Complementary
Prompt Sampling (CPS), a heuristic sampling strategy that exploits these
predictions to improve generation quality. Switching between placements
across nearby edits can introduce audible artifacts, so CPS also applies
local prompt consistency.

Similar to token revision in discrete diffusion models trained with
uniform noise~\cite{vonrutte2025gidd}, we find that \method{} can also
edit plausible speech token sequences, although its training sources
are initialized with random tokens from \(p_0\). We use this
generalization for end-to-end editing without external edit-region
localization. However, this capability is not explicitly trained,
creating a training--inference mismatch
~\cite{yao2026selfgenerated}. To mitigate this mismatch and correct
errors in model-generated speech, we continue Edit Flow sampling after
generation as post-generation refinement. With the Edit Flow model trained on 10K h of GigaSpeech, experimental
results demonstrate competitive performance in both zero-shot TTS and
speech editing.

\begin{figure*}[t]
    \centering
\includegraphics[width=0.98\textwidth,trim=3 145 0 52,clip]{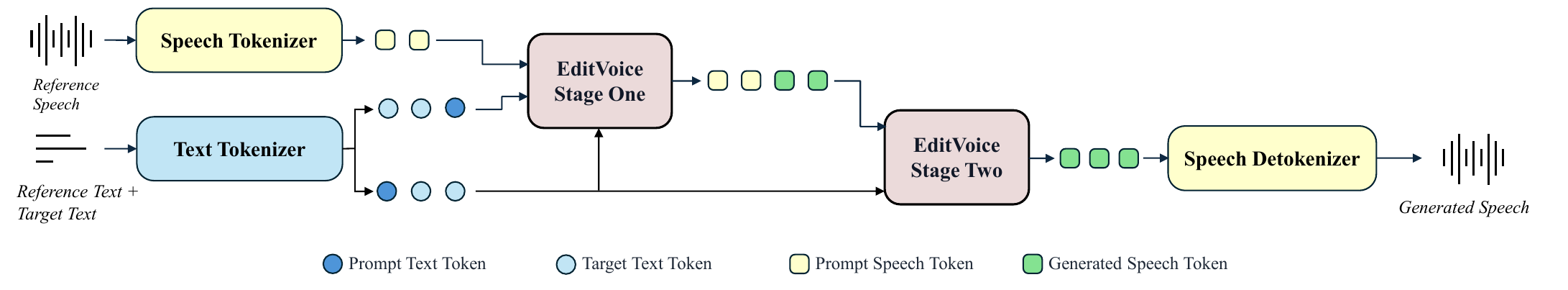}
    \caption{Overview of the \method{} zero-shot TTS inference pipeline.
Stage one generates semantic speech tokens using CPS, and stage two
applies post-generation refinement to the generated sequence before
speech detokenization.}
    \label{fig:overview}
\end{figure*}

\section{Background: Edit Flows}
\label{sec:background}

\subsection{Variable-Length Edit Process}
\label{sec:editflow_process}

Edit Flows~\cite{havasi2025edit} formulate non-autoregressive
generation directly on the variable-length sequence space
\begin{equation}
\mathcal X=\bigcup_{N=0}^{N_{\max}}\mathcal V^N,
\label{eq:sequence_space}
\end{equation}
where \(\mathcal V\) is a discrete vocabulary. A continuous-time
Markov chain (CTMC) on \(\mathcal X\) transitions between sequences
through insertion, deletion, and substitution operations. Writing
\(u_t^\theta(\omega\mid x_t)\) for the instantaneous rate of edit
\(\omega\), insertion and substitution are factorized into an
operation rate \(\lambda\) and a normalized token distribution \(q\):
\begin{equation}
\begin{aligned}
u_t^\theta(\mathrm{ins}_{i,a}\mid x_t)
&=\lambda^{\mathrm{ins}}_{t,i}\,q^{\mathrm{ins}}_{t,i}(a),\\
u_t^\theta(\mathrm{sub}_{i,a}\mid x_t)
&=\lambda^{\mathrm{sub}}_{t,i}\,q^{\mathrm{sub}}_{t,i}(a),\\
u_t^\theta(\mathrm{del}_{i}\mid x_t)
&=\lambda^{\mathrm{del}}_{t,i},
\end{aligned}
\label{eq:rate_factor}
\end{equation}
where the dependence of \(\lambda\) and \(q\) on \(x_t\) and \(t\)
is omitted for brevity. A rate \(\lambda\) over a sampling step \(h\)
corresponds to the finite-step event probability
\(p=1-\exp(-h\lambda)\).

\subsection{Training with Auxiliary Alignments}
\label{sec:editflow_training}

Directly constructing a conditional CTMC between two variable-length
sequences is generally intractable because multiple edit sequences may
connect the same endpoints. Edit Flows introduce an auxiliary aligned
space \(\mathcal Z=(\mathcal V\cup\{\varepsilon\})^M\), where
\(\varepsilon\) is an auxiliary blank symbol, together with a mapping
\(f_{\mathrm{rm}}:\mathcal Z\rightarrow\mathcal X\) that removes blanks. For aligned endpoints \(z_0,z_1\in\mathcal Z\), the conditional path factorizes across aligned coordinates:
\begin{equation}
\begin{aligned}
p_t(z_t\mid z_0,z_1)
&=\prod_{j=1}^{M}\Big[(1-\kappa_t)\delta_{z_0^j}(z_t^j)
+\kappa_t\delta_{z_1^j}(z_t^j)\Big],\\
X_t&=f_{\mathrm{rm}}(Z_t),
\end{aligned}
\label{eq:conditional_path}
\end{equation}
where \(\kappa_0=0\) and \(\kappa_1=1\). An unresolved aligned
coordinate corresponds to one insertion, deletion, or substitution in
\(\mathcal X\). Following Edit Flows~\cite{havasi2025edit}, the
training objective is
\begin{equation}
\begin{aligned}
\mathcal L_{\mathrm{EF}}=\mathbb E\Bigg[{}
&\sum_{\omega\in\Omega(X_t)}u_t^\theta(\omega\mid X_t)\\[-0.2em]
&-\beta_t\!\sum_{j:z_t^j\neq z_1^j}
\log u_t^\theta(\omega_j\mid X_t)\Bigg],
\end{aligned}
\label{eq:editflow_loss}
\end{equation}
where \(\beta_t=\dot\kappa_t/(1-\kappa_t)\),
\(\Omega(X_t)\) is the set of valid edits, and \(\omega_j\) denotes
the edit corresponding to replacing \(z_t^j\) with \(z_1^j\).
\section{EditVoice}
\label{sec:method}

\subsection{Speech Infilling with Edit Flows}
\label{sec:infilling}

Figure~\ref{fig:overview} summarizes the two-stage zero-shot TTS
inference pipeline.
\method{} models semantic speech token sequences with Edit Flows,
conditioned on text and speech context.

\textbf{Training.}
Given speech tokens \(x_1\) and transcript \(y\), we replace a random
span with \(L_0=50\) tokens sampled from the empirical distribution
\(p_0\), retaining the rest as acoustic context. Following Edit
Flows~\cite{havasi2025edit}, we align the sampled span and the original
target span using auxiliary blanks \(\varepsilon\). In the aligned
pair, \(z_0^j=\varepsilon\) requires an insertion,
\(z_1^j=\varepsilon\) requires a deletion, and unequal nonblank tokens
require a substitution. This allows a fixed initial length to be paired with target spans of
varying lengths. We then use the conditional path and training
objective in Eqs.~\eqref{eq:conditional_path} and
\eqref{eq:editflow_loss}. Random span placement exposes context on
either side of the generated region.

\emph{Zero-shot TTS.}
Let \(s_p,y_p,\hat y\) denote the prompt speech tokens, prompt
transcript, and target text. We initialize the target with \(L_0\)
samples from \(p_0\) and place the fixed prompt before or after it:
\(([s_p;x_t],[y_p;\hat y])\) or
\(([x_t;s_p],[\hat y;y_p])\). Edits are restricted to the target state, while insertions and
deletions allow its length to vary during sampling. The output semantic
tokens are decoded by the frozen CosyVoice~2 acoustic decoder
~\cite{du2024cosyvoice2}.

\emph{Localized cascade speech editing.}
A text diff and word-level forced alignment localize the region changed
from transcript \(y\) to \(y'\). We replace that region in source speech
\(x^{\mathrm{src}}\) with \(L_0\) samples from \(p_0\), keep the
remainder fixed, and disable edits outside the region. During acoustic
decoding, source Mel frames are copied for retained tokens, while the
edited region is regenerated by the CFM and vocoder.

\subsection{Complementary Prompt Sampling}
\label{sec:cps}

Speech-infilling training makes both prefix and suffix prompt placements
available for zero-shot TTS and speech editing. We find that
the two placements can produce complementary operation-rate
predictions: an edit assigned a low rate by one placement may receive
a high rate from the other. Switching between placements across
successive local edits can also introduce inconsistencies and audible
artifacts. We therefore introduce Complementary Prompt Sampling (CPS),
a heuristic sampling strategy that combines rate-guided prompt selection
with local prompt consistency.

Given the current target state \(x_t\), we evaluate the same model with
the prompt speech \(s_p\) placed before and after \(x_t\):
\begin{equation}
\mathcal P_{\mathrm{pre}}
=
([s_p;x_t],[y_p;\hat y]),
\qquad
\mathcal P_{\mathrm{suf}}
=
([x_t;s_p],[\hat y;y_p]).
\label{eq:placements}
\end{equation}
For brevity, we omit the sampling-time and position indices \(t,i\)
below. Each placement \(r\in\{\mathrm{pre},\mathrm{suf}\}\) predicts
operation rates \(\lambda_r^o\) for
\(o\in\{\mathrm{ins},\mathrm{del},\mathrm{sub}\}\) and token
distributions \(q_r^o\) for \(o\in\{\mathrm{ins},\mathrm{sub}\}\).

\begin{figure}[!t]
    \centering
\includegraphics[width=\linewidth,trim=189 25 432 22,clip]{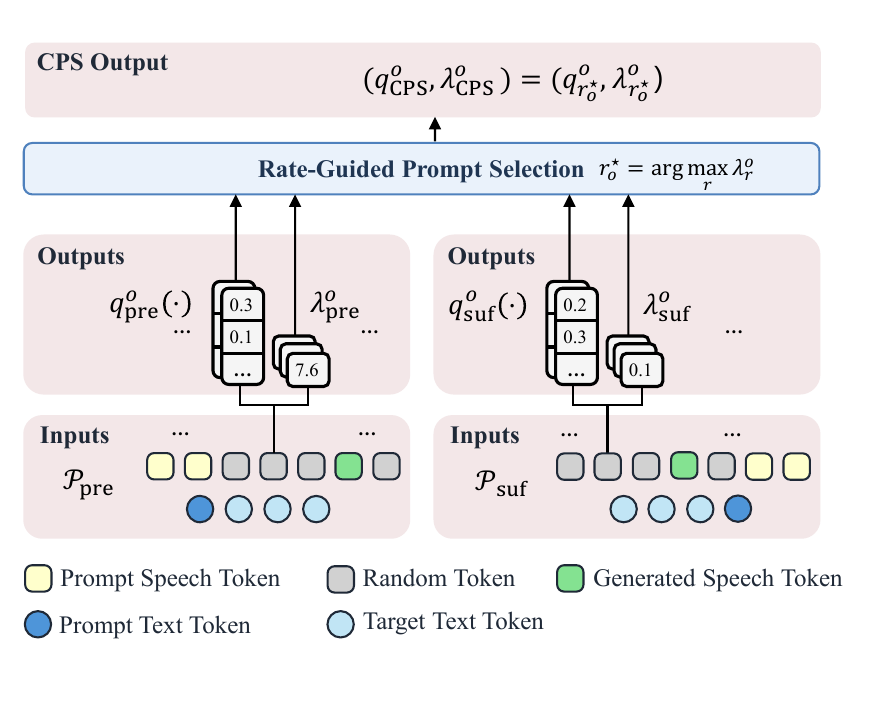}
    \caption{Illustration of Complementary Prompt Sampling (CPS).
The same target state is evaluated with prefix and suffix prompt
placements. For each edit operation, rate-guided prompt selection
chooses the placement assigning the larger operation rate and uses its
corresponding token distribution. Local prompt consistency is omitted
for clarity.}
    \label{fig:cps}
\end{figure}

\textbf{Rate-guided prompt selection.}
For each operation
\(o\in\{\mathrm{ins},\mathrm{del},\mathrm{sub}\}\), CPS selects the
prompt placement with the larger operation rate:
\begin{equation}
r_o^\star
=
\arg\max_{r\in\{\mathrm{pre},\mathrm{suf}\}}
\lambda_r^o,
\qquad
\lambda_{\mathrm{CPS}}^o
=
\lambda_{r_o^\star}^o.
\label{eq:cps_select}
\end{equation}
For insertion and substitution, the token distribution is taken from
the same placement:
\begin{equation}
q_{\mathrm{CPS}}^o
=
q_{r_o^\star}^o,
\qquad
o\in\{\mathrm{ins},\mathrm{sub}\}.
\label{eq:cps_token}
\end{equation}
We record the selected prompt placement for each inserted or substituted
token.

\textbf{Local prompt consistency.}
Rate-guided selection may switch between prompt placements across
successive edits in the same local region, which can produce
inconsistent token predictions and audible artifacts. A conflict occurs
when a later deletion or substitution of an inserted or substituted
token selects the other prompt placement.

For \(t>\tau\), once a conflict is detected, subsequent edits to the
conflicted token and its immediate neighbors, up to one token on each
side, use the prompt placement recorded for the conflicted token.
Tokens generated from this region inherit the same placement. Delaying
this rule until \(t>\tau\) allows early sampling to more fully exploit
the complementary rate predictions from both placements.

\begin{table*}[!t]
\centering
\caption{Zero-shot TTS results. For \method{}, 10K h denotes the
GigaSpeech data used to train the Edit Flow model.
$\dagger$ denotes systems using the same S3Tokenizer2 tokenizer and
CosyVoice~2 detokenizer as \method{}; bold and underline denote the
best and second-best synthesized results. T2S RTF measures only
text-to-semantic-token generation.}
\label{tab:tts_main}
\renewcommand{\arraystretch}{0.87}
\setlength{\tabcolsep}{1.1pt}
\begin{tabular}{@{}lccccccccc@{\hspace{4pt}}c@{\hspace{4pt}}c@{}}
\toprule
\smash{\raisebox{-1.2ex}{Model}}
& \smash{\raisebox{-1.2ex}{Params.}}
& \smash{\raisebox{-1.2ex}{Train (h)}}
& \multicolumn{3}{c}{Seed-TTS EN}
& \multicolumn{3}{c}{LibriSpeech-PC}
& \smash{\raisebox{-1.2ex}{NMOS $\uparrow$}}
& \smash{\raisebox{-1.2ex}{SMOS $\uparrow$}}
& \smash{\raisebox{-1.2ex}{T2S RTF $\downarrow$}}\\
\cmidrule(lr){4-6}\cmidrule(lr){7-9}
& & &
WER $\downarrow$ & SIM-o $\uparrow$ & UTMOS $\uparrow$
& WER $\downarrow$ & SIM-o $\uparrow$ & UTMOS $\uparrow$
& & & \\
\midrule
Ground Truth & -- & -- 
& -- & -- & --
& -- & -- & --
& 3.77 & 4.11 & --\\

\midrule
\multicolumn{12}{l}{\emph{Autoregressive}}\\

CosyVoice2$^{\dagger}$~\cite{du2024cosyvoice2}
& 0.5B & 167K
& 2.58 & 0.659 & \textbf{4.17}
& 1.80 & 0.655 & \underline{4.39}
& \underline{3.88} & 3.84 & 0.5313\\

CosyVoice3~\cite{du2025cosyvoice3}
& 0.5B & 1M
& 2.00 & 0.697 & 3.97
& 1.77 & 0.695 & 4.30
& -- & -- & --\\

VoxCPM~\cite{zhou2025voxcpm}
& 0.5B & 1.8M
& 1.86 & \textbf{0.729} & 3.82
& 1.92 & \textbf{0.719} & 4.21
& -- & -- & --\\

Qwen3-TTS-12Hz (Base)~\cite{hu2026qwen3tts}
& 0.6B & 5M
& 1.63 & 0.708 & \underline{4.13}
& 1.66 & \underline{0.700} & \textbf{4.41}
& -- & -- & --\\

\midrule
\multicolumn{12}{l}{\emph{Non-autoregressive}}\\

OmniVoice-Emilia~\cite{zhu2026omnivoice}
& 0.6B & 100K
& 1.63 & \underline{0.715} & 3.88
& \textbf{1.61} & 0.696 & 4.23
& \textbf{3.92} & \textbf{4.12} & --\\

F5-TTS~\cite{chen2025f5tts}
& 0.3B & 100K
& 1.75 & 0.671 & 3.69
& 2.12 & 0.653 & 3.90
& -- & -- & --\\

ZipVoice (16 NFE)~\cite{zhu2025zipvoice}
& 0.1B & 100K
& 1.59 & 0.695 & 3.82
& 1.76 & 0.671 & 3.99
& -- & -- & --\\

MaskGCT~\cite{wang2024maskgct}
& 1.0B & 100K
& 2.64 & 0.713 & 3.56
& 2.39 & 0.692 & 3.91
& -- & -- & \underline{0.2791}\\

\midrule
\method{}$^{\dagger}$ (64 NFE)
& 0.4B & 10K
& \textbf{1.48} & 0.658 & \underline{4.13}
& \underline{1.63} & 0.663 & 4.37
& 3.85 & \underline{3.88} & --\\

\method{}$^{\dagger}$ (32 NFE)
& 0.4B & 10K
& \underline{1.53} & 0.659 & 4.12
& 1.73 & 0.662 & 4.37
& -- & -- & --\\

\method{}$^{\dagger}$ (16 NFE)
& 0.4B & 10K
& 1.55 & 0.659 & 4.11
& 2.11 & 0.660 & 4.36
& -- & -- & \textbf{0.0989}\\

\bottomrule
\end{tabular}
\end{table*}

\subsection{End-to-End Editing and Post-Generation Refinement}
\label{sec:editing_refine}
Although training sources are initialized with random tokens sampled
from \(p_0\), we find that \method{} can also edit plausible speech
token sequences. We use this generalization for end-to-end editing and
post-generation refinement.

\textbf{End-to-end editing.}
Given source speech, its transcript, and a target transcript, we
initialize the editable state with the complete source speech and
additionally use the same speech as a fixed prompt for speaker
conditioning. The model is conditioned on the concatenated source and
target transcripts and directly determines which source tokens to edit
and which to retain, without external edit-region localization.
Acoustic decoding follows the same retention procedure as localized
editing.

\textbf{Post-generation refinement.}
Prior work on discrete diffusion shows that generated samples can be
improved by post-generation self-correction~\cite{vonrutte2025gidd}.
We find that \method{} can also revise model-generated tokens, although
such tokens are not used as training sources. We therefore keep the
final sampling time fixed and continue Edit Flow sampling for several
steps using only the prefix prompt placement to correct residual content
errors.

\section{Experiments}
\label{sec:experiments}

\subsection{Experimental Setup}
\label{sec:setup}

\textbf{Model and training.}
Our 444.7M-parameter Edit Flow model uses a 14-layer bidirectional
LLaMA-style Transformer (1024 hidden, 4096 FFN, 16 attention heads,
4 KV heads), a 6-layer Conformer text encoder, and frozen S3Tokenizer2
speech tokens. We train the Edit Flow model on 10K h of GigaSpeech for 230K steps
with learning rate \(5\times10^{-4}\) and 8K warmup. We use the linear
scheduler \(\kappa_t=t\), and conditions are jointly dropped with
probability 0.1 for CFG.   

Zero-shot TTS uses the frozen CosyVoice~2 acoustic decoder. To better
handle noisy source speech in editing, we use Kimi-Audio
BigVGAN~\cite{kimiteam2025kimiaudio} and fine-tune the CosyVoice~2 CFM
on 350 h of GigaSpeech to match its Mel configuration.

\textbf{Inference.}
For zero-shot TTS, the 64/32/16-NFE settings use 56/24/8 CPS
generation steps, respectively, followed by 8 refinement steps.
Additional CFG and CPS evaluations within each sampling step are not
counted separately in NFE. We use nucleus sampling (\(p=0.7\)),
\(T(t)=1-0.2t\), na\"ive rate CFG~\cite{havasi2025edit} with strength 4,
and \(\tau=0.3\).
Speech editing uses 24 CPS generation steps followed by 8 refinement
steps (32 NFE total), with CFG strength 6. For all tasks, refinement
uses only the prefix prompt placement with CPS disabled. For end-to-end
editing, retained Mel segments shorter than six frames are regenerated
to avoid unnatural artifacts.

\textbf{Evaluation.}
For zero-shot TTS, we evaluate on Seed-TTS Eval EN and
LibriSpeech-PC test-clean using WER, SIM-o, and UTMOS. For
intelligibility, WER is measured between the ASR transcription of
synthesized speech and the input text. We use Whisper-large-v3
~\cite{radford2023robust} for Seed-TTS Eval EN and a HuBERT-based ASR
model~\cite{hsu2021hubert} for LibriSpeech-PC test-clean. Speaker
similarity is measured by SIM-o using a WavLM-based
ECAPA-TDNN~\cite{chen2022wavlm,desplanques2020ecapa}, and speech
quality by UTMOS~\cite{saeki2022utmos}. Subjective evaluation reports
naturalness mean opinion score (NMOS) and speaker similarity MOS
(SMOS) from 20 participants. T2S RTF is measured on a single 32-GB
NVIDIA V100 GPU and includes only text-to-semantic-token generation. For direct NAR comparisons, we use models
trained on at most 100K h, while larger-scale AR systems are included
as broader references.

For speech editing, we evaluate on RealEdit~\cite{peng2024voicecraft}.
We use Whisper-large-v3~\cite{radford2023robust} for WER,
WavLM-TDNN~\cite{chen2022wavlm} for SIM-o,
UTMOS~\cite{saeki2022utmos} for speech quality, and
MOSNet~\cite{lo2019mosnet}. For cascade editing, edit regions are
localized by MFA~\cite{mcauliffe2017mfa}, with each baseline using its
original extension setting and \method{} using 0.12 s on both sides. For \method{}, we also report the fraction of source Mel frames directly
preserved in the edited speech for both cascade and end-to-end editing,
computed as \(\sum_i R_i/\sum_i N_i\), where \(N_i\) and \(R_i\) are
the numbers of source and preserved Mel frames, respectively.

All baseline results are obtained from official checkpoints using the
evaluation models described above.

\subsection{Zero-Shot TTS and Speech Editing}
\label{sec:main_results}

\textbf{Zero-shot TTS.}
Table~\ref{tab:tts_main} shows that \method{} achieves the lowest WER
on Seed-TTS Eval EN and the second-lowest WER on LibriSpeech-PC.
Compared with the AR CosyVoice~2 baseline, which uses the same
S3Tokenizer2 tokenizer and CosyVoice~2 detokenizer, \method{} reduces
WER from 2.58 to 1.48 on Seed-TTS Eval EN and from 1.80 to 1.63 on
LibriSpeech-PC. It also improves speaker similarity on LibriSpeech-PC,
with SIM-o increasing from 0.655 to 0.663.
These results show that variable-length Edit Flow generation can retain
the duration flexibility of AR decoding while providing stronger content
consistency. The 16-NFE variant reaches a T2S RTF of 0.0989, compared
with 0.2791 for MaskGCT and 0.5313 for CosyVoice~2.

\begin{table}[!t]
  \centering
  \caption{Objective RealEdit results; bold and underline denote the
  best and second-best results within each setting.}
  \label{tab:realedit}
  \renewcommand{\arraystretch}{0.87}
  \setlength{\tabcolsep}{1.7pt}
  \begin{tabular}{lccccc}
  \toprule
  Model & Type & WER $\downarrow$ & SIM-o $\uparrow$
  & MOSN $\uparrow$ & UTMOS $\uparrow$\\
  \midrule
  Ground Truth & -- & 5.59 & -- & 3.22 & 3.40\\
  \midrule
  \multicolumn{6}{l}{\emph{Cascade editing}}\\
  FluentSpeech~\cite{jiang2023fluentspeech}
  & NAR & 5.80 & 0.949 & 3.18 & 2.68\\
  VoiceCraft~\cite{peng2024voicecraft}
  & AR & 5.89 & 0.973 & 3.14 & 3.34\\
  SSR-Speech~\cite{wang2024ssrspeech}
  & AR & \underline{4.75} & \textbf{0.986}
  & \underline{3.20} & \textbf{3.36}\\
  \method{}
  & NAR & \textbf{4.40} & \underline{0.981}
  & \textbf{3.22} & \underline{3.35}\\
  \midrule
  \multicolumn{6}{l}{\emph{End-to-end editing}}\\
  Ming-UniAudio~\cite{yan2025minguniaudio}
  & AR & 10.40 & 0.969
  & \underline{3.24} & 3.22\\
  CosyEdit~\cite{chen2026cosyedit}
  & AR & \textbf{3.94} & \underline{0.973}
  & 3.23 & \underline{3.24}\\
  \method{} E2E
  & NAR & \underline{4.67} & \textbf{0.975}
  & \textbf{3.26} & \textbf{3.50}\\
  \bottomrule
  \end{tabular}
\end{table}

\textbf{Speech editing.}
Table~\ref{tab:realedit} shows that, in cascade editing, \method{}
achieves the best WER and MOSNet. Compared with the AR SSR-Speech
baseline, the NAR \method{} reduces WER from 4.75 to 4.40. In
end-to-end editing, \method{} achieves the best SIM-o, MOSNet, and
UTMOS and the second-best WER.

Without external localization, end-to-end \method{} leaves part of the
source sequence unchanged and copies the corresponding Mel frames,
preserving 31.5\% of the source Mel frames. In contrast, the compared
E2E systems regenerate the complete edited utterance. This is consistent
with the highest SIM-o of 0.975. MFA-localized cascade editing preserves
76.7\%, leaving substantial room to improve source preservation in
end-to-end editing.

\subsection{Ablation and Analysis}
\label{sec:analysis}

\begin{table}[!t]
\centering
\leftalignedcaption{CPS and refinement ablation on Seed-TTS Eval EN.}{tab:ablation}
\renewcommand{\arraystretch}{0.87}
\setlength{\tabcolsep}{4.0pt}
\begin{tabular}{lccc}
\toprule
Inference & WER $\downarrow$ & SIM-o $\uparrow$ & UTMOS $\uparrow$\\
\midrule
\multicolumn{4}{l}{\emph{56 generation NFE}}\\
Prefix
& 1.65\% & 0.653 & 4.08\\
Suffix
& 1.66\% & 0.656 & 4.08\\
CPS w/o Prompt Consistency
& 1.61\% & \underline{0.658} & 4.08\\
CPS
& \underline{1.54\%} & \textbf{0.659} & \underline{4.11}\\
\midrule
\multicolumn{4}{l}{\emph{56 generation + 8 refinement NFE}}\\
Prefix + Refinement
& 1.58\% & 0.653 & 4.10\\
CPS + Refinement
& \textbf{1.48\%} & \underline{0.658} & \textbf{4.13}\\
\bottomrule
\end{tabular}
\end{table}

\textbf{Effect of CPS.}
Table~\ref{tab:ablation} shows similar performance for the prefix and
suffix prompt placements. Combining their complementary predictions
without local prompt consistency reduces WER from 1.65--1.66\% to
1.61\%. Local prompt consistency further reduces WER to 1.54\% and
improves UTMOS from 4.08 to 4.11, consistent with the audible artifacts
caused by frequent prompt switching.

\begin{figure}[!t]
    \centering
\includegraphics[width=0.90\columnwidth]{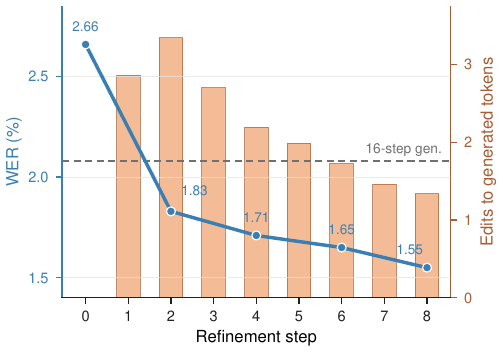}
    \caption{Post-generation refinement on Seed-TTS Eval EN.
Starting from the same 8-step CPS output, WER and edits to generated
tokens are measured over 8 refinement steps. The dashed line denotes
16-step CPS generation without refinement.}
   \label{fig:refinement}
\end{figure}

\textbf{Refinement analysis.}
Figure~\ref{fig:refinement} shows that \method{} continues to delete or
substitute model-generated tokens after generation, although such tokens
are not used as training sources. WER decreases from 2.66\% to 1.55\%
over refinement, and 8 generation steps plus 2 refinement steps already
outperform 16 generation steps without refinement.

\section{Conclusion}
\label{sec:conclusion}

We introduced \method{}, a variable-length NAR zero-shot TTS model
based on Edit Flows. Speech-infilling training supports localized
editing and two prompt placements, whose complementary predictions are
combined by CPS. The learned editing capability also generalizes to
plausible speech token sequences, enabling end-to-end editing and
training-free refinement. Experiments demonstrate competitive
zero-shot TTS and speech-editing performance with efficient NAR
generation.

\clearpage
{\centering\bfseries ACKNOWLEDGMENTS\par}
\smallskip
No funding was received for conducting this study. The authors have no
relevant financial or nonfinancial interests to disclose.
\medskip

{\centering\bfseries COMPLIANCE WITH ETHICAL STANDARDS\par}
\smallskip
This study was performed in line with the principles of the Declaration of
Helsinki. Approval was granted by the Ethics Committee of Xiamen University.
All participants in the subjective listening tests provided informed consent.
Only anonymized ratings were collected.
\medskip

\bibliographystyle{IEEEbib}
\bibliography{references_arxiv}

\end{document}